\documentstyle[prd,aps,amsmath,twocolumn,amsfonts]{revtex}
\begin{document}
\newcommand{\bra}{\langle}
\newcommand{\ket}{\rangle}
\newcommand{\tbf}[1]{{\bf #1}}
\newcommand{\tit}[1]{{\it #1}}
\newcommand{\cl}[1]{{\cal #1}}
\newcommand{\half}{{\scriptstyle \frac12}}
\newcommand{\Rcite}[1]{ref.\cite{#1}}
\newcommand{\intRR}{\int\limits_{-\infty}^{\infty}}
\newcommand{\intR}{\int\limits_{0}^{\infty}}
\newcommand{\LRD}[1]{\frac{{{\displaystyle\leftrightarrow}
\atop {\displaystyle\partial}}}{\partial #1}}
\newcommand{\lrd}[1]{\stackrel{\displaystyle \leftrightarrow}
{\displaystyle\partial_{#1}}}
\newcommand{\diff}[1]{\partial /{\partial #1}}
\newcommand{\Diff}[1]{\frac{\partial}{\partial #1}}
\newcommand{\DiffM}[2]{\frac{\partial #1}{\partial #2}}
\newcommand{\DiffT}[1]{\frac{\partial^2}{\partial {#1}^2}}
\newcommand{\hc}[1]{#1^{\dagger}}
\newcommand{\ds}{\displaystyle}

\newcount\secnum \secnum=0
\def\newsec{\advance\secnum by 1
\vskip 0.5cm
{\tbf{\the\secnum.\quad}}\nobreak}

\title{Unruh quantization in presence of a condensate}
\author{V.A. Belinskii\thanks{E-mail: belinski@icra.it}}
\address{INFN and ICRA, Rome University "La Sapienza",00185 Rome, Italy,}
\address{and IHES, F-91440 Bures-sur-Yvette, France}
\author{N.B. Narozhny\thanks{E-mail: narozhny@theor.mephi.ru},
A.M. Fedotov\thanks{E-mail: a\_fedotov@yahoo.com}, and V.D. Mur,}
\address{Moscow Engineering Physics Institute, 115409 Moscow, Russia}
\maketitle

\begin{abstract}
We have shown that the Unruh quantization scheme can be realized
in Minkowski spacetime in the presence of Bose-Einstein condensate
containing infinite average number of particles in the zero boost
mode and located basically inside the light cone. Unlike the case
of an empty Minkowski spacetime  the condensate provides the
boundary conditions necessary for the Fulling quantization of the
part of the field restricted only to the Rindler wedge of
Minkowski spacetime.

\end{abstract}


The Unruh effect \cite {Davies,Unruh} consists of the statement
that any detector uniformly accelerated in the empty Minkowski
space (MS) reveals a universal response as if it were emersed into
a thermal bath with the temperature $T_{DU}=g/2\pi$, where $g$ is
the proper acceleration of the detector\footnote{We use natural
units $c=\hbar=k_B=1$}. The universality of the detector response
is based on interpretation of the Minkowski vacuum state in terms
of Fulling-Unruh particles states which appear in the Unruh
quantization scheme \cite{Unruh} alternative to the standard plane
wave quantization (see also Ref. \cite{BD} and citations therein).
However, we have shown in our earlier paper \cite{PRD}, see also
Refs. \cite{PhLet,Ann}, that the Unruh construction \cite{Unruh}
was not a valid quantization scheme in the empty MS. This
conclusion stems from the fact that the Unruh scheme neglects the
contribution of the singular zero boost mode to the total field
amplitude.

The purpose of this paper is to present a situation for which the
Unruh quantization procedure could be realized. We argue that this
is the case for MS which is not empty but filled with a background
of special type. We hope that our consideration will clarify the
drawbacks of the Unruh approach to the problem in empty MS. For
the sake of simplicity we will restrict our consideration to
two-dimensional case.

Let us consider a coherent state for a massive neutral scalar
field
\begin{equation}
\label{BE_condensate}
|\,coh\,\ket=Z_{coh}^{-1/2}\,\exp\left\{\intRR
\frac{dp}{\sqrt{2\pi\epsilon_p}}\;f(p)\; \hc{a_p}\right\}|0\ket_M,
\end{equation}
where $\hc{a_p}$ is a conventional creation operator of a particle
in the plane wave state with momentum $p$ and energy
$\epsilon_p=\sqrt{p^2+m^2}$, $f(p)$ is a complex weight function,
$Z_{coh}=e^{\bar N}$ is the normalization constant and $\bar N$ is
the expectation value of total number of particles.

The momentum distribution of particles in the state
(\ref{BE_condensate}) is described by
\begin{equation}
\label{p_distr} d\bar N_p=\bra \,coh\,|\,\hc{a_p}a_p\,|\,coh\,\ket
dp=\frac{|f(p)\,|^2}{2\pi\epsilon_p}\,dp\;,
\end{equation}
while the expectation values $\bar N$ of the total number of
particles and of the field energy $\bar E$ in the state
(\ref{BE_condensate}) are given by Eqs. (\ref{bar N}, \ref{Av(E)})
respectively
\begin{equation}
\label{bar N} \bar N=\intRR d\bar
N_p=\intRR|f(p)\,|^2\frac{dp}{2\pi\epsilon_p}\;,
\end{equation}
\begin{equation}\label{Av(E)}
\bar E=\intRR \epsilon_p\,d\bar
N_p=\intRR|f(p)\,|^2\frac{dp}{2\pi}\;.
\end{equation}
Clearly, the total number of particles and the field total energy
in the state (\ref{BE_condensate}) are infinite if the function
$|f(p)\,|^2$ decreases too slowly at large momenta. In particular,
the integrals in Eqs. (\ref{bar N}, \ref{Av(E)}) diverge for the
case $f(p)=c=$ const.

When dealing with the Unruh problem it is convenient to quantize
the field in the basis of eigenfunctions $\Psi_{\kappa}(x),\,
x=(t,z)$, of the Lorentz boost operator rather than in the
plane-wave basis \cite{Unruh,PRD}. These functions have the
following integral representation, see, e.g., \cite{PRD}
\begin{equation}
\label{Boost_Modes} \Psi_{\kappa}(x)=\frac{1}{2^{3/2}\pi}\intRR
dq\,e^{im(z\sinh{q}-t\cosh{q})-i\kappa q}\;,
\end{equation}
and boost destruction operators $b_\kappa$ can be expressed in
terms of plane wave operators $a_p$ as
\begin{equation}
\label{b} b_{\kappa}=\intRR
\frac{dp}{\sqrt{2\pi\epsilon_p}}\left(\frac{\epsilon_p+p}{m}\right)
^{i\kappa}a_p\;.
\end{equation}
The two quantization schemes are unitary equivalent \cite{PRD}.

Let us introduce the notation $|\,c\,\ket$ for the coherent state
(\ref{BE_condensate}) with $f(p)=c=$ const. Using Eq. (\ref{b}) it
can be easily represented in terms of boost operators as
\begin{equation}
\label{BE_cond1} |\,c\,\ket=Z_c^{-1/2}\,e^{c \hc{b_0}}|0\ket_M,
\end{equation}
where $\hc{b_0}$ is the creation operator of a particle in the
state corresponding to the singular zero boost mode $\Psi_0(x)$,
\begin{equation}
\label{Wh f} \Psi_0 (x)=-i\sqrt{2}\Delta^{(+)}(x;m)\;,
\end{equation}
with $\Delta^{(+)}(x;m)$ being the positive-frequency Whightman
function, see Ref. \cite{PRD}. Thus, in the framework of the boost
quantization scheme the state (\ref{BE_condensate}) with $f(p)=$
const contains an infinite number of particles concentrated in a
single quantum state, namely in the state with zero value of the
boost quantum number $\kappa$. Hence, in a sense, this state is
equivalent to the Bose-Einstein condensate. It is well known that
the Whightman function [which determines the zero mode Eq.
(\ref{Wh f})] vanishes exponentially fast for large spacelike
distances away from the light cone. It is clear therefore that
particles of the condensate are located basically inside the light
cone. Moreover, the energy density in this state is infinite at
the light cone. This is closely related to singular properties of
the zero boost mode. Indeed, near the light cone $x_{+}x_{-}\to 0$
($x_{\pm}=t\pm z$) the zero mode $\Psi_0(x)$ behaves \cite{PRD} as
\begin{equation}
\label{delta lc} \Psi_0(x)\sim -\frac{1}{2\sqrt{2}\pi}\ln[-m^2
x_+x_-+i\varepsilon\,{\rm sgn}\,(t)]\;,
\end{equation}
where we assume that $\varepsilon\rightarrow +0$ and ${\rm
sgn}\,(t)=0,$ if $t=0.$ Therefore even in a one-particle state
$|B_0\ket\propto \hc{b_0}|0_M\ket$ the expectation value of energy
density $T^{00}$ possesses nonintegrable singularities at the
cone, $\bra B_0|T^{00}|B_0\ket\sim (1/8\pi^2)(x_{\pm})^{-2},\quad
x_{\pm}\to 0$. The zero value of the normalization constant
$Z_c^{-1/2}$ in Eq.~(\ref{BE_cond1}) reflects the well-known fact
that the state of condensate belongs to representation of
canonical commutation relations (CCR) unitary inequivalent to the
conventional Fock representation (see, e.g., \cite{Umez}).

It is worth noting that the state $|\,c\,\ket$ essentially differs
from the standard condensate commonly used in condensed matter
physics. Since the boost generator does not commute with the
Hamiltonian, there is no definite value of energy for any state of
the field with a fixed value of $\kappa$ including $\kappa=0$.
Therefore the state of the field with $\kappa=0$ is not the ground
state of the system, and moreover, it is not stationary. Hence the
only common feature of the state $|\,c\,\ket$ and the standard
condensate is the infinite number of particles concentrated in a
single quantum state. It is clear that the boost condensate
$|\,c\,\ket$ could not come into existence for a system of free
bosons like it happens in standard situation and should be
considered as {\it initially} prepared.

It follows from the definition (\ref{BE_cond1}) that the state
$|\,c\,\ket$ satisfies the relation
\begin{equation}\label{c_p_st}
b_{\kappa}|\,c\,\ket=c\delta(\kappa)|\,c\,\ket.
\end{equation}
Let us introduce new operators $\widetilde{b}_{\kappa}$ by
\begin{equation}\label{B_shift}
b_{\kappa}=\widetilde{b}_{\kappa}+c\delta(\kappa).
\end{equation}
These operators satisfy the usual CCR:
\begin{equation}\label{CCR1}
[\widetilde{b}_{\kappa},\hc{\widetilde{b}}_{\kappa'}]=
\delta(\kappa-\kappa'),\quad
[\widetilde{b}_{\kappa},\widetilde{b}_{\kappa'}]=
[\hc{\widetilde{b}}_{\kappa},\hc{\widetilde{b}}_{\kappa'}]=0.
\end{equation}
The transformation (\ref{B_shift}) is often called the boson shift
\cite{Umez}. Let us call the quasiparticles associated with the
shifted destruction and creation operators
$\widetilde{b}_{\kappa}$, $\hc{\widetilde{b}}_{\kappa}$ boostons.
Since
\begin{equation}{\label{b_vac}}
\widetilde{b}_{\kappa} |\,c\,\ket=0,
\end{equation}
the zero mode condensate $|\,c\,\ket$ can be considered a vacuum
state with respect to boostons (of course, it is not vacuum with
respect to conventional particles). Hence the boostons can be
considered as excitations of the pure condensate state. Such
situation is ordinary for condensed matter physics.

Now let us show that it is possible to perform Unruh quantization
in the presence of the zero mode condensate. We start with
decomposition of the field operator in terms of the boost modes
\begin{equation}
\label{Quant_BM} \phi_M(x)=\intRR
d\kappa\{b_{\kappa}\Psi_{\kappa}(x)+ \text{h.c.}\},
\end{equation}
If we apply the boson shift (\ref{B_shift}) to
Eq.~(\ref{Quant_BM}), the free field $\phi_M(x)$ acquires  the
form
\begin{equation}\label{B_shift1}
\phi_M(x)=\widetilde{\phi}_M(x)+\varphi_c(x),
\end{equation}
where $\varphi_c(x)=c\Psi_0(x)+\text{h.c.}$ is the classical part
of the field (the condensate), and
\begin{equation}
\label{Quant_BM1} \widetilde{\phi}_M(x)=\intRR
d\kappa\{\widetilde{b}_{\kappa} \Psi_{\kappa}(x)+\text{h.c.}\},
\end{equation}
is the quantized booston field. $c$-number function $\varphi_c(x)$
in Eq. (\ref{B_shift1}) has the meaning of the condensate average
of the field $\phi_M(x)$
$$\varphi_c(x)=\bra\,c\,|\phi_M(x)|\,c\,\ket,$$
and according to terminology of Ref. \cite{Umez} forms an
"extended macroscopical object". So, in the presence of the
condensate the field $\phi_M(x)$ describes a system of
non-interacting boostons and an extended macroscopical object,
compare Ref. \cite{Umez}.

The presence of the extended macroscopical object $\varphi_c(x)$
in Eq. (\ref{B_shift1}) allows us to omit the contribution of the
zero mode (\ref{Wh f}) in expansion (\ref{Quant_BM1}). This means
that the booston field can be represented in the form
\begin{equation}
\label{Pv}\widetilde{\phi}_M(x)\,\approx\,\text{P.v.}\intRR
d\kappa\, \{\widetilde{b}_{\kappa}\Psi_{\kappa}(x)+\text{h.c.}\}.
\end{equation}
Physically this approximation means neglecting quantum
fluctuations of the zero mode field against the background of the
infinite number of particles in the condensate. This can be
illustrated symbolically by the following argument. It follows
from the commutation relations (\ref{CCR1}) that
$[\widetilde{b}_0,\hc{\widetilde{b}_0}]=\delta(\kappa)\vert_{\kappa=0}$,
so that if $c\ne 0$ then $\widetilde{b}_0\sim \sqrt{\delta(\kappa)
\vert_{\kappa=0}}\ll c\cdot\delta(\kappa)\vert_{\kappa=0}$.
Formally, this estimate should be understood in a weak sense. This
means that it is valid for all matrix elements of the operator
$\widetilde{b}_0$ between the states with finite number of
boostons. The matrix elements are supposed to be smeared with
respect to the spectral parameter $\kappa$.

Following Unruh \cite{Unruh}, we can now split the interval of
integration over $\kappa$ in the RHS of Eq.(\ref{Pv}) into two for
$\kappa=\mu>0$ and $\kappa=-\mu<0$ and after simple transformation
arrive to the following form for the field decomposition
\begin{equation}
\label{Quant_Uc} \phi_M(x)=\intR
d\mu\,\{\widetilde{r}_{\mu}R_{\mu}(x)+
\widetilde{l}_{\mu}L_{\mu}^*(x)+\text{h.c.}\}+ \varphi_c(x).
\end{equation}
Here $R_{\mu}(x)$ and $L_{\mu}(x)$ are the Unruh modes
\cite{Unruh,PRD} and $\widetilde{r}_{\mu}$ and
$\widetilde{l}_{\mu}$ are defined as
\begin{eqnarray}\nonumber
\widetilde{r}_{\mu}=\frac1{\sqrt{2\sinh\pi\mu}}
\left\{e^{\pi\mu/2}\widetilde{b}_{\mu}+e^{-\pi\mu/2}
\hc{\widetilde{b}_{-\mu}}\right\},
\\ \label{rl_qp_Ops}
\quad \widetilde{l}_{\mu}=\frac1{\sqrt{2\sinh\pi\mu}}
\left\{e^{\pi\mu/2}\widetilde{b}_{-\mu}+e^{-\pi\mu/2}
\hc{\widetilde{b}_{\mu}}\right\}.
\end{eqnarray}

The relations (\ref{rl_qp_Ops}) are equivalent to the Bogolubov
transformation. They define new operators $\widetilde{r}_{\mu},
\widetilde{l}_{\mu}$ which act in the Fock space
${\mathcal{H}}(\widetilde{r}_{\mu}, \widetilde{l}_{\mu})$ unitary
inequivalent to the Fock space
${\mathcal{H}}(\widetilde{b}_{\kappa})$. In
${\mathcal{H}}(\widetilde{r}_{\mu}, \widetilde{l}_{\mu})$ the
operators $\widetilde{r}_{\mu}, \widetilde{l}_{\mu}$ and their
Hermitian conjugates have the sense of destruction, creation
operators for $\widetilde{r}$ and $\widetilde{l}$ Unruh
quasiparticles which we will call rightons and leftons
respectively. The vacuum state from
${\mathcal{H}}(\widetilde{b}_{\kappa})$ (booston vacuum, or
condensate state $|\,c\,\ket$) can be now represented as
superposition of $n$-particle states of rightons and leftons
\cite{Umez,Unruh,PRD}
\begin{multline}
\label{c_cont} |c\,\ket=Z^{-\frac12}\sum\limits_{n=0}^{\infty}
\int\limits_{0}^{\infty}d\mu_{1}\ldots\int\limits_{0}^{\infty}
d\mu_{n}\,\exp\left(-\pi\sum\limits_{i=1}^{n}\mu_i\right)\\ \\
\times|\widetilde{1}_{\mu_1},\ldots\widetilde{1}_{\mu_n}\ket_{L}\otimes
|\widetilde{1}_{\mu_1},\ldots\widetilde{1}_{\mu_n}\ket_{R}.
\end{multline}
The constant $Z$ in (\ref{c_cont}) is of course infinite but again
this only reflects unitary inequivalence of two Fock spaces. We
should emphasize that Eq. (\ref{c_cont}) is valid only if we are
allowed to neglect the contribution of the zero mode to the
booston field $\widetilde{\phi}_M(x)$. This can be done in MS only
in the presence of the condensate and is absolutely inadmissible
in empty MS. In the latter case the set of Unruh modes is
incomplete, so that the decomposition (\ref{Quant_Uc}) for the
field in empty MS $\,$($\varphi_c(x)=0$) simply does not exist and
the operators (\ref{rl_qp_Ops}) have no sense of destruction
operators for any particles in MS, see Ref. \cite{PRD}. Hence the
Eq. (\ref{c_cont}) loses its sense if we change the state
$|\,c\,\ket$ by $|0_M\ket$.

Though Fock spaces ${\mathcal{H}}(\widetilde{r}_{\mu},
\widetilde{l}_{\mu})$ and ${\mathcal{H}}(\widetilde{b}_{\kappa})$
are unitary inequivalent, action of the operators
$\widetilde{r}_{\mu}, \widetilde{l}_{\mu}$ on the state vectors
from ${\mathcal{H}}(\widetilde{b}_{\kappa})$ are well defined by
Eqs. (\ref{rl_qp_Ops}) and we can easily derive the Unruh formula
\cite{Unruh,BD}
\begin{equation}
\label{rr_av} \bra
c|\hc{\widetilde{r}_{\mu}}\widetilde{r}_{\mu'}|c\ket=
(e^{2\pi\mu}-1)^{-1}\,\delta(\mu-\mu').
\end{equation}
Now, following Unruh \cite{Unruh} we will try to interpret this
formula from the point of view of a Rindler observer.

It is highly important that owing to the absence of the zero mode
in decomposition (\ref{Pv}) the field $\widetilde{\phi}_M(x)$
automatically satisfies the condition
\begin{equation}\label{0_cond}
\widetilde{\phi}_M(0,0)=0, \end{equation} which follows from the
relation $\Psi_{\kappa}(0,0)=(1/\sqrt{2})\,\delta(\kappa)$
\cite{PRD} and should be understood in weak sense. Note that in
the presence of the condensate the field $\phi_M(x)$, see Eq.
(\ref{B_shift1}), obeys the same condition at the origin
$\phi_M(0,0)=\infty$ as the classical field $\varphi_c(x)$. We
have asserted in Ref. \cite{PRD} that, due to translation
invariance, the relation of the type (\ref{0_cond}) for quantum
field in empty MS would mean that all matrix elements of the
operator $\phi_M(x)$ between physically realizable states are
identically equal to zero. Now this assertion does not work since
translation invariance of MS is broken in presence of the
nonuniform extended macroscopical object, the condensate.

Let us now consider the field $\widetilde{\phi}_M(x)$ Eq.
(\ref{Pv}) restricted to the Rindler wedge of MS,
$\,x=(\eta,\rho), \; t=\rho\sinh\eta\,, \;z=\rho\cosh\eta\,,\;
\rho\,>\,0\,,\,-\infty<\eta<\infty\,.$ This field as a function of
Rindler coordinates satisfies the Klein-Fock-Gordon equation and
obeys the boundary condition $\widetilde{\phi}_M(0,\eta)=0$.
Hence, as shown in Ref. \cite{PRD}, it can be quantized in
accordance with the Fulling procedure \cite{Fulling} which leads
to the concept of Fulling quasiparticles. We call here Fulling
quanta quasiparticles rather than particles because the classical
background $\varphi_c(\rho,\eta)$ is not equal to zero, so that
the Fulling quanta are excitation over the condensate in the
Rindler wedge. In view of boundary condition (\ref{0_cond}) it
follows from Eqs. (5.16) of Ref. \cite{PRD} that the destruction
operator for the Fulling quasiparticle coincides with the
destruction operator $\widetilde{r}_{\mu}$ for righton. Therefore
the factor in front of $\delta$-function in Eq. (\ref{rr_av}) can
be interpreted as the average number of Fulling quasiparticles
seen by an observer moving in the Rindler wedge. Moreover, since
$\mu=\omega/g$ and $\eta=g\tau$ where $\omega$ is the energy of
Fulling quantum and $\tau$ is the proper time of a uniformly
accelerated observer in MS (world line of which lies entirely in
Rindler wedge), we arrive to the following conclusion. Any
detector moving with a constant proper acceleration in MS filled
with zero-mode condensate $|\,c\,\ket$ responds as if it had been
immersed into a thermal bath of Fulling quasiparticles at the
Davies - Unruh temperature $T_{DU}=g/2\pi$.

This statement dramatically differs from the conventional
formulation of the Unruh effect \cite{Unruh,BD} since the latter
assumes acceleration of the detector in the empty MS. The main
physical difference between these two situations is that the
quantum dynamical degree of freedom associated with zero boost
mode, which is an obstacle for validity of Unruh quantization in
vacuum, can be naturally ignored in the background singular at the
light cone, i.e. when particles occupying the zero mode quantum
state form an extended macroscopical object. It is important that
neglecting of zero mode contribution to the field amplitude
automatically leads to zero boundary condition for the booston
field restricted to the Rindler wedge, thus providing a
possibility for Fulling quantization.

Certainly concentration of infinite number of particles in a
single quantum state is an idealization. Physically, when talking
about the zero mode condensate, we should understand not a
particular mode but a narrow (with respect to $\kappa$) wave
packet with the center at $\kappa=0$. It can be seen from the
integral representation (\ref{Boost_Modes}) that at $t=0$ such a
wave packet will be narrow with respect to $z$ also. At $t\neq 0$
the wave packet remains narrow in the sense that it has sharp (but
not infinite) maxima at the surface of the light cone,
$x_{\pm}=0$. It is clear that expectation value of energy density
in this quantum state is not singular any more, compare with the
text after Eq. (\ref{delta lc}). One should keep in mind of course
that the wave packet should remain narrow during the time of
observation, i.e. the time $\Delta\tau$ which one substitutes
instead of $\delta$-function at $\mu=\mu'$ in Eq. (\ref{rr_av})
according to the rule $\delta(0)=\Delta\tau/2\pi$. The infinite
number of particles in the wave packet state should be understood
as macroscopically large. The Unruh construction in presence of
such condensate is an approximation of course, but evidently the
accuracy of this approximation looks very similar to the accuracy
of the well known thermodynamic limit in the many body problems.

We are very grateful to our colleagues  A.V. Berkov, S.R. Kelner,
Yu.E. Lozovik, B.N. Narozhny and D.N. Voskresensky for helpful
discussions. This work was supported by the Russian Fund for Basic
Research and by Ministry of Education of Russian Federation.

\end{document}